\documentclass{article}
\usepackage{graphicx}
\usepackage[a4paper, total={6in, 8in}]{geometry}
\usepackage{placeins}
\usepackage{url}

\title{Mathematical modelling to inform outbreak response vaccination}
\author{Manjari Shankar$^{1,*}$, Anna-Maria Hartner$^{1,2}$, 
\\ Callum R.K. Arnold $^3$,  Ezra Gayawan$^4$, Hyolim Kang$^5$, Jong-Hoon Kim$^6$, Gemma Nedjati Gilani$^1$, 
\\ Anne Cori$^1$, Han Fu$^5$, Mark Jit$^{5,11}$, Rudzani Muloiwa$^7$, Allison Portnoy$^{8, 9}$, 
\\ Caroline Trotter$^{1,10}$, Katy A M Gaythorpe$^1$}

\begin{document}

\maketitle
$^1$ Medical Research Council Centre for Global Infectious Disease Analysis, Imperial College London, London, UK
$^2$ Centre for Artificial Intelligence in Public Health Research, Robert Koch Institute, Wildau, Germany
$^3$ Center for Infectious Disease Dynamics, Pennsylvania State University, University Park, PA 16802, USA
$^4$ Department of Statistics, Federal University of Technology, Akure, Nigeria
$^5$ Department of Infectious Disease Epidemiology, Faculty of Epidemiology and Population Health, London School of Hygiene \&\ Tropical Medicine, London, UK
$^6$ Department of Epidemiology, Public Health, Impact, International Vaccine Institute, Seoul, South Korea
$^7$ Department of Paediatrics, Faculty of Health Sciences, University of Cape Town, Groote Schuur Hospital, Cape Town, South Africa
$^8$ Department of Global Health, Boston University School of Public Health, Boston, United States
$^9$ Center for Health Decision Science, Harvard T.H. Chan School of Public Health, Boston, United States.
$^{10}$ Department of Veterinary Medicine and Pathology, University of Cambridge, Cambridge, UK
$^{11}$ School of Public Health, University of Hong Kong, Hong Kong Special Administrative Region, China

\textit{$^*$ Corresponding author: Manjari Shankar, m.shankar@imperial.ac.uk}

\section*{Abstract}

Mathematical models are established tools to assist in outbreak response. They help characterise complex patterns in disease spread, simulate control options to assist public health authorities in decision-making, and longer-term operational and financial planning. In the context of vaccine-preventable diseases (VPDs), vaccines are one of the most-cost effective outbreak response interventions, with the potential to avert significant morbidity and mortality through timely delivery. Models can contribute to the design of vaccine response by investigating the importance of timeliness, identifying high-risk areas, prioritising the use 
 of limited vaccine supply, highlighting surveillance gaps and reporting, and determining the short- and long-term benefits. In this review, we examine how models have been used to inform vaccine response for 10 VPDs, and provide additional insights into the challenges of outbreak response modelling, such as data gaps, key vaccine-specific considerations, and communication between modellers and stakeholders. We illustrate that while models are key to policy-oriented outbreak vaccine response, they can only be as good as the surveillance data that inform them.

\textbf{Keywords}: vaccination, impact, outbreak, immunisation, mathematical modelling, vaccine

\newpage
\section{Background}

Vaccine-preventable diseases (VPDs) continue to pose a significant global health challenge. Often attributed to gaps in vaccination coverage, the emergence and spread of outbreaks of VPDs are characterised by a disproportionately high burden in low and middle income countries (LMICs) \cite{amoako2022preponderance}. Almost 103 countries have seen measles outbreaks in the last 5 years due to low vaccine coverage, demonstrating the urgency of closing such immunisation gaps and protecting those at-risk \cite{unicef2024}. Limited access to clean water and sanitation has additionally resulted in an acute resurgence in cholera outbreaks across 23 countries this year, increasing demand for vaccines from the emergency global stockpiles \cite{gtfcc2024}. According to the IA2030 scorecard \cite{ia20302022}, of the 40 known outbreaks in 2022 that had an outbreak response vaccination strategy, only 18\% of these had a timely detection and response, emphasising the need to improve health system responses to decrease burden of disease.

Of the several effective outbreak response interventions, vaccines are among the most cost-effective, and rapidly aid containment and reduce mortality and morbidity \cite{Cairns2011Should}, \cite{who2020IA2030, Trovato2020Viral, carter2023modeling}.  Since 2000, the implementation of outbreak response immunisation programs in LMICs has averted 5.81M cases and saved 327k lives across 210 outbreaks of 4 vaccine-preventable diseases  \cite{delport2024estimating}. During an outbreak, however, complex patterns in disease spread\cite{metcalf2015seven, beraud2018mathematical} and uncertainties in epidemiological and operational parameters \cite{li2019concurrent, metcalf2015seven, van2012epidemic} can hinder the optimal design of outbreak response vaccination strategies. Given these complexities, the immediate use of mathematical models can help project the effect of vaccine deployment strategies \cite{metcalf2015seven, mcbryde2020role} and assess their sustainability based on key considerations such as vaccine availability, at-risk populations, competing health system priorities and long-term financial and operational implications \cite{guttieres2023modeling}. Such models can be used to rapidly test key hypotheses, estimate available parameters, evaluate past interventions and project the impact of future strategies to inform public health policy, \cite{zelner2022rapid, metcalf2020mathematical,Louz2010Emergence}. 

The insights from model-based approaches can contribute to national and global policy recommendations on the timing and impact of vaccination strategies, while accounting for variable input data and assumptions \cite{recker2016assessing}. Thus, despite several challenges around the availability of suitable data, spatial and social heterogeneity in risk and incidence, and communication between modellers and policymakers in the event of an outbreak \cite{beraud2018mathematical, recker2016assessing}, mathematical models remain valuable tools in evaluating vaccination impact.

Previous studies have examined interactions between modelling and policy in defining outbreak response as a part of specially commissioned research groups \cite{ferguson2020report,wu2020nowcasting, brooks2021modelling} or for specific diseases or geographies \cite{De2009Controlling, lessler2016impact, graham2019measles, mcbryde2020role, james2021use, hadley2024modelling} . However, a consolidated overview of how mathematical modelling can assist outbreak response vaccination across all vaccine-preventable diseases (VPDs) is lacking. This review synthesises study findings and the collective experience of modellers to demonstrate how mathematical models have informed various aspects of an outbreak response vaccination strategy and led to their extensive use for contextual policy guidance. The focus is specifically on the modelling of outbreaks across 10 VPDs \footnote{\textit{Typhoid*, Dengue*, MenA*, yellow fever*, Measles*, Cholera*, COVID-19*, Ebola, Chikungunya, Mpox. Those with an * are modelled in the Vaccine Impact Modelling Consortium.}} where a vaccine is currently available for use in outbreak response. Recognising data uncertainties, we discuss key criteria such as the definition of an outbreak, and the data required to arrive at a robust model alongside vaccine-specific considerations for modelling outbreak response. Finally, we touch on the limitations of modelling vaccine use during an outbreak and explore key considerations for communication.

\section{Main text}

\subsection{Significance of modelling in addressing vaccine policy questions}

Mathematical models are useful tools to synthesise available data and influence vaccine policy across different phases of an outbreak. To understand the significance of iterative policy-oriented modelling, it is helpful for the purposes of this review to classify outbreak response distinctly into the investigative, scale-up and control phases \cite{morgan2019decision}. These phases are illustrated below in figure \ref{fig:outbreak_timeline}. 

\begin{figure}[htbp]
    \centering
    \includegraphics[scale=0.2]{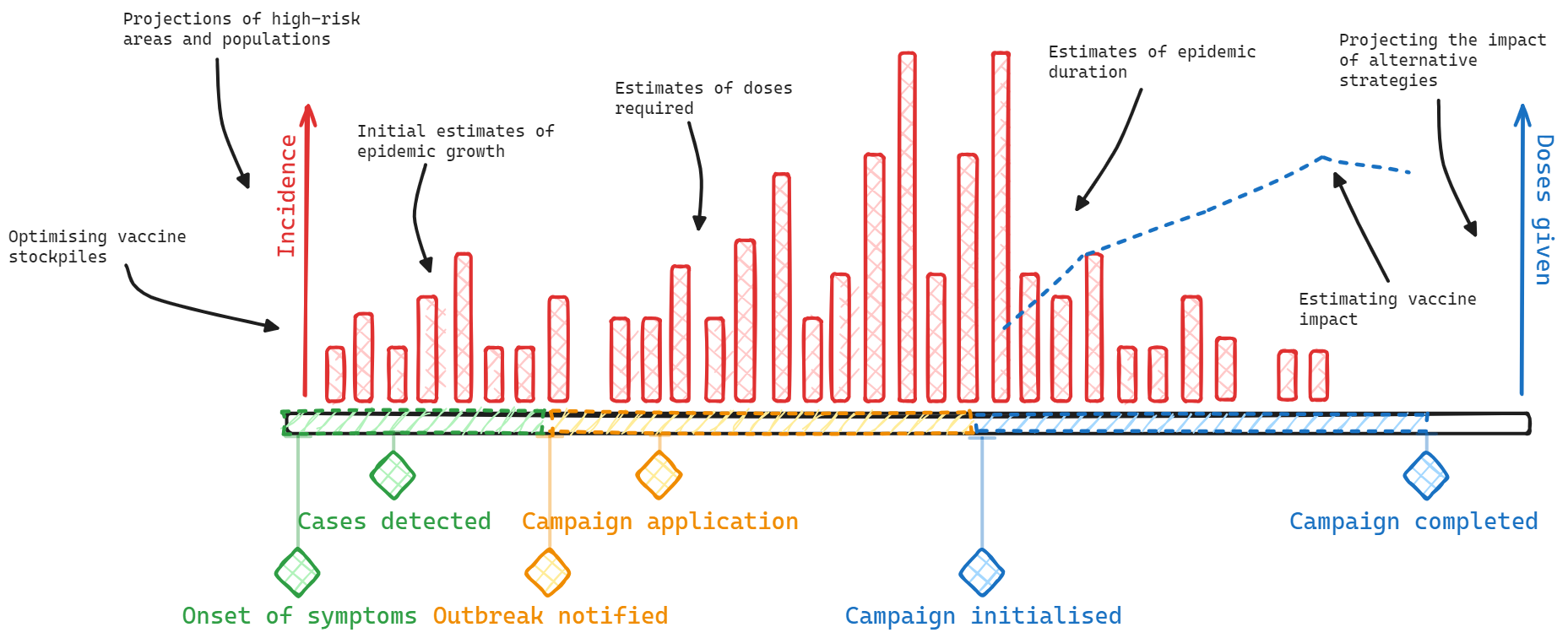}
    \caption{Timeline of outbreak included detection and outbreak response vaccination campaign. Red vertical bars indicate incidence or similar. Blue dotted lines indicate vaccination coverage or doses given. Diamonds indicate key points along the timeline and colour indicates phases where green = investigative phase, yellow = scale-up phase, blue = control phase. Black text indicates potential modelling outputs at each stage.}
    \label{fig:outbreak_timeline}
\end{figure}

The earliest stage of the outbreak requires surveillance or detection followed by rapid collection of data; modelling at this stage can provide early insights into transmission dynamics and the immediate impact of the outbreak. Vital statistics such as the characteristics of the pathogen, disease burden, transmission rate, population at-risk, and demand for healthcare can be difficult to obtain in a timely and consistent way and may not be directly observable, particularly in the early stages of an outbreak. However, it may be possible to synthesise evidence from previous outbreaks of the same pathogen to better define the parameter space.  In scaling up an outbreak response, modelling can account for heterogeneity in the population to tailor vaccine interventions and prioritise accordingly, subject to data availability Models can also be used to draft control strategies, to identify gaps in surveillance and reporting, to estimate the actual need for vaccines as well as help with the prioritisation and stockpiling for future use. Modelling can also be of assistance outside the outbreak timeline, either identifying at-risk areas before an outbreak or reviewing intervention effectiveness following an outbreak. The impact of an outbreak response vaccination program depends on factors such as the timeliness of the response at each stage alongside the rapid identification of target population, estimating vaccine availability and optimising health system capacity. 

As an example, a real-time modelling exercise by Graham et al. \cite{graham2019measles} to respond to a measles outbreak with a catch-up vaccination campaign in Guinea amidst the Ebola crisis demonstrated the usefulness of model-based projections of population risk and future incidence on priority setting and planning. Furthermore, modelling has been used to estimate critical values such as the severity of disease or case fatality ratio which are vital for healthcare provision \cite{servadio2021estimating,verity2020estimates,garske2017heterogeneities}. This integration of mathematical modelling into policy design has provided critical insights into outbreak dynamics and the effectiveness of responses  \cite{overton2020using}. 

\subsubsection{Outbreak response timeliness}
When assessing and responding to an outbreak, swift action is required. 
This applies to both understanding the real-time situation and to implementing interventions. 
Numerous outbreaks have seen models used to project real-time incidence and burden, and Ebola is a key example. 
Across different outbreaks, Ebola incidence has been projected given intervention scenarios, including the 2014-2016 Guinea epidemic \cite{ajelli2016spatiotemporal}, 2018 Equateur, DRC epidemic \cite{kelly2019projections}, and the 2018-2020 outbreak in North Kivu and Ituri Provinces, DRC \cite{worden2019projections}.
Similarly, the benefits of rapid outbreak responses have been quantified across multiple modelling studies for a range of diseases in the context of: logistical and operational constraints \cite{tao2018logistical}, alert and action thresholds for responding to outbreaks \cite{trotter2015response, ferrari2014time,cooper2019spatiotemporal}, and alternative scenarios around outbreak response timing \cite{zhao2018modelling}.
Throughout these studies, the same qualitative conclusion has appeared- that more rapid outbreak response vaccinations provides better results, however, modelling allows this conclusion to be \emph{quantified} for the context and outbreak in question.
For example, a study on the 2015/6 outbreak of yellow fever in Angola found that a 60-day delay in vaccine deployment would have more than doubled the observed deaths and a delay of 180 days would have led to a five-fold increase in deaths \cite{zhao2018modelling}.

\subsubsection{Identifying high-risk areas}
Effective prevention and control of infectious disease outbreaks requires the consideration of heterogeneity in disease risk, incidence and effectiveness of vaccine interventions . During an outbreak, geostatistical models can help situate socioeconomic mobilisation and public health decision-making through characterising spatial dynamics and optimising available interventions like vaccines \cite{franch2020spatial, Gross_2020,marion2022modelling}. They can estimate burden at different geographical scales and suggest areas of higher risk thereby facilitating the effective deployment of vaccines or other public health measures. These types of models have been utilised for outbreaks of COVID \cite{grauer2020strategic}, Ebola \cite{wells2019ebola, wells2019exacerbation, bellan2015statistical} and polio \cite{Voorman2021Real-time} to suggest optimised vaccination strategies.

The risk of outbreaks may be spatially heterogeneous and influenced by environmental, climatic or landscape-related factors \cite{verdonschot2014flight,krol2024landscape}.  This is evident in YF as well as chikungunya and dengue amongst others \cite{KANG2024488,cattarino2020mapping,lim2023systematic}. Understanding these factors can aid in preventative measures and help with ensuring surveillance is available in at-risk locations, therefore informing the investigative phase of outbreak response. For example, multiple studies have examined the potential of spread of YF from endemic areas to vulnerable populations, in part prompted by the exportation of cases from an outbreak in Angola in 2016 to China \cite{codecco2004risk,johansson2014whole,sakamoto2018modeling, brent2018international,dorigatti2017international,cracknell2021yellow} and more recently focused on Djibouti, Somalia and Yemen \cite{fraser2024yellow} to inform surveillance in potential and moderate risk countries.

\subsubsection{Prioritisation of limited vaccines and optimising stockpiles}
Vaccination is one tool in the suite of outbreak response activities and, depending on the pathogen, it may be the primary form of response, such as for yellow fever. 
However, ensuring that sufficient doses are available at the right time, in the right place, requires careful planning ahead of time.  
As a result, stockpiles are developed that can be deployed to tackle outbreaks at short notice. 
The size, location and timing of production for vaccine stockpiles, as well as the optimal deployment approach vary by context. 
Mathematical modelling is one way of optimising the size and deployment of vaccine dose stockpiles. 
For example, the stockpile of yellow fever (YF) vaccines is limited and new doses take one year to produce \cite{yfinfograph}. But, mathematical models have been used to show fractional dosing can be safely utilised to stretch supplies when necessary and prevent outbreaks as the population-wide benefits of higher coverage overshadow the potential loss in efficacy for an individual \cite{chen2020modelling, wu2016fractional}.

\subsubsection{Modelling as a tool for highlighting surveillance gaps and reporting}
By synthesising data and evidence, mathematical models can be used to identify areas of greater uncertainty or influence. They can also estimate the under-reporting of incidence in an outbreak; epidemics of meningitis have occurred in the African meningitis belt for more than 100 years but, whilst the largest \emph{reported} epidemic occurred in 1996, it is likely the true incidence was almost double that reported as routine reporting systems faltered and families avoided seeking healthcare \cite{greenwood1999manson, mohammed2000severe}.
Similarly, YF has a non-specific symptom set and this can affect reporting.  Modelling has been used to estimate the severity spectrum based on historic outbreaks \cite{johansson2014whole, servadio2021estimating}, the probability that a case may be reported \cite{gaythorpe2021global}, or to project burden in areas where surveillance data is absent \cite{shearer2018existing, gaythorpe2021global}.

\subsubsection{Considering multiple vaccines and interventions}
Mathematical modelling can be used to assess relative benefits of interventions and approaches and this can include, although less commonly, multiple vaccines or interventions. Often further outcomes are examined such as cost-effectiveness or healthcare burden rather than the more common indicators of mortality such as reported deaths. For example, in the case of Ebola, some studies have moved away from purely epidemiological modelling to understand the cost-effectiveness and pricing of vaccines \cite{obeng2021economic,obeng2022ee58,guttieres2023modeling}. 
Examining multiple pathogens and interventions in the same modelling framework can lead to informative results on which interventions are universally optimal, vs just effective for one pathogen. For example, when chikungunya, dengue, Zika and yellow fever were considered together, the usage of insecticide and insecticide-treated bed-nets was found to be optimal irrespective of which diseases were included \cite{claypool2022assessing}.
Similarly, weighing the relative benefits of interventions for the same disease has been discussed, for example including YF in the Expanded Program on Immunization (EPI) for Nigeria was found to be more cost-effective than emergency response \cite{monath1993should}.

\subsection{Key model inputs}
At each phase of the outbreak, there are key data that can inform modelling and/or decision making; this includes information on the outbreak response itself. 
Figure \ref{fig:outbreak_timeline} illustrates an example outbreak timeline with key notification points as well as the potential modelling that can take place at different phases.

\subsubsection{Defining an outbreak}
The definition of an outbreak varies by context and pathogen, and in some cases, over time. Understanding the criteria for the beginning and end of an outbreak allows modellers and public health officials to assess the outbreak magnitude, duration, and severity, thus informing a proportionate response. Modelling has been used to assess the confidence that an outbreak is over based on time since final reported case \cite{djaafara2021quantitative}.

Classifying the beginning and end of an outbreak, as well as whether reported incidence is endemic or epidemic, is critical for producing realistic and actionable model outputs.  Brady and colleagues \cite{brady2015dengue} test approximately 102 variable outbreak definitions on a dataset of reported dengue cases in Brazil to show that inconsistency in these can hinder an effective outbreak response and establish the need for clear quantitative definitions to support modelling exercises. In the case of yellow fever, one reported case constitutes an outbreak, so understanding the under-reporting and reporting delays are key to understanding when transmission may have occurred \cite{mondiale2017eliminate}. In some cases, an outbreak is defined by a period where the effective reproductive number is above 1, the epidemic threshold \cite{jombart2021real, stolerman2023using, teklehaimanot2004alert, leclere2017automated}. In such cases, it may be possible to define an automated threshold for detection to improve response timeliness  \cite{stern1999automated, salmon2016monitoring, shmueli2010statistical, unkel2012statistical, farrington2003outbreak}.

\subsubsection{Pathway to outbreak detection}
In practice, it is often not possible to observe the transmission of infection events that lead to an outbreak, only the change in the reported burden \cite{farrington2003outbreak}.
This highlights the importance of capturing uncertainty at each stage of the outbreak modelling. 
For example, the speed and accuracy of diagnostic tests (if they are available) should be considered when developing alerts or thresholds for outbreak detection, as well as background-noise infections (non-target diseases that present with clinically similar symptoms).
Médecins Sans Frontières use different measles outbreak definitions based on whether there is IgM confirmation, as well as the recency and coverage of vaccination campaigns \cite{msf2024guidelines}. 
Model simulations of the underlying dynamics and testing components can be used to explore the interaction between diagnostic test uncertainty, levels of background noise, testing rates, and outbreak and alert definitions providing insight into appropriate outbreak thresholds and response triggers \cite{brett2018anticipating, brett2020detecting}. 
Further, modelling methods to account for delayed and reduced reporting rates have been developed, but due to their computational complexity they may not be feasible to deploy in real-time and/or resource-constrained environments that are typical of outbreak settings \cite{gostic2020practical}. 

\subsubsection{Data requirements for modelling}
As seen during the recent COVID-19 pandemic, challenges in finding and accessing data and its varying quality and coverage has underscored the need for a better data ecosystem for modelling needs in the future \cite{shadbolt2022challenges}. 
Despite this, modellers and the COVID-19 response benefited from analytical and visualisation capabilities and collective efforts to improve models \cite{chen2022rampvis}.
Using locally available, granular data alongside country-owned modelling has formed the basis of user-friendly tools for outbreak response \cite{mandal2022imperfect}.
This approach improves both the socialisation of model outputs as well as the quality of the model itself through the integration of relevant data.
Key data sources for epidemiological modelling of outbreak response vaccination include aspects such as case counts, disease occurrence, seroprevalence surveys and historical outbreak response timing \cite{lim2023systematic, KANG2024488}. 
Other information such as demography, mobility and historic immunisation coverage are also critical to establish the epidemiological state of the population at the time of the outbreak.
As noted later, the quality of modelled outputs is contingent on the quality of input data and assumptions. 

\subsection{Vaccine-specific considerations in modelling outbreaks}

\subsubsection{Common and unique vaccine questions}
Vaccine-preventable disease outbreaks can present unique, disease-specific questions, but there are often common analysis needs that are relevant for many epidemics- particularly around healthcare demand forecasting or timing of interventions. 
For yellow fever, vaccine-specific considerations often include the time required to manufacture the vaccine due to frequent supply shortages; similarly, this often results in the need for fractional dosing during outbreaks \cite{wu_2016}. 
For yellow fever, Dengue, Ebola, and Mpox, there are challenges in our understanding of immune correlates of protection \cite{domingo_2019, guttieres_2023, king_2015, recker_2016}; ongoing discussions for yellow fever consider whether booster-doses are needed or if assumptions of lifelong protection are appropriate \cite{wu_2016}. 
For Dengue, there are differences in the variations of efficacies in endemic settings and across different serotypes \cite{recker_2016}. 
Additionally, differences have been seen in efficacies between naive individuals and individuals with dengue antibodies \cite{recker_2016}. 
For Mpox, there are large uncertainties on effectiveness that must be taken into account, as current research assumes it confers similar protection to smallpox \cite{yuan_2022}. 
For Ebola and measles, the duration of vaccine-induced protection is unknown, though for the latter the timescale is greater than a single outbreak \cite{guttieres_2023, king_2015}. 
The vaccines for Ebola Virus Disease also have additional considerations, as supply constraints often mean there is a trade-off between priority geographies, policy aims, and feasibility; strategies like ring vaccination may not always be possible \cite{guttieres_2023, king_2015, henao2017efficacy}.
Recent studies have focused on assessing the use cases of the novel vaccines in a variety of settings with varying model structures \cite{henao2017efficacy, chen2021hybrid, bodine2017potential, potluri2022model, xie2019data, wells2015harnessing, kucharski2016effectiveness}. 
And as seen with COVID-19, future vaccine considerations may need to consider the possibility of immune escape, as this could jeopardise vaccine-induced herd immunity \cite{caldwell_2021}.

\subsubsection{Evaluating long and short-term benefits}
During outbreaks, policymakers often rely on modelling estimates for both short and long-term decision making. Short-term timelines often focus on the emergency aspects of the response — guiding policy and potential actions \cite{baker_2022}. Later in an outbreak, long-term decision making may involve dealing with competing objectives or other social and economic costs \cite{baker_2022}. Vaccination activities, whether they are outbreak response campaigns or routine immunisation, can also have both immediate and longer-term benefits. During an outbreak-response, the aim of the vaccination activity is usually to stop the spread of an outbreak thus reducing the burden of severe disease and deaths. However, depending on the pathogen and vaccine, such activities can have benefits over the lifetimes of vaccinees that should not be overlooked in impact assessments. This can be captured by different views of vaccine impact such as by calendar year, for more immediate effects, or by vaccinated birth cohort, to capture longer-term benefits \cite{echeverria2021can}. However, it is also important to consider the time window that an intervention is evaluated over which can be linked to how the end of an outbreak is declared \cite{djaafara2021quantitative}.

\subsection{Key considerations for communication}
Ideally, local, within-country, and context-specific capacity for modelling and relationships between stakeholders and modellers should already be established in advance of an outbreak; this ensures decision science can move at the pace required to prevent disease transmission and deal with ongoing uncertainty \cite{baker_2022}. 	Currently, however, several countries lack the technical capacity, relationships, or communication skills for modelling evidence to be used effectively in outbreak situations \cite{mbachu_2024, si_2022}. The barriers to the use of modelling evidence by policymakers are varied. Most frequently, policymakers cite a lack of relevant research, i.e. models do not address the concerns or situations policymakers face to be useful in decision making \cite{oliver_2014, levin_2024, freebairn_2018, lee_2010, si_2022}. In situations where models do not yet exist, policymakers note there is no time or opportunity to use the research evidence \cite{oliver_2014, mbachu_2024, grieve_2023, si_2022} or find barriers to the cost of model development \cite{oliver_2014, si_2022}. Further, policymakers and other users have stated that they are unable to understand and interpret the evidence modellers provide \cite{oliver_2014, mbachu_2024, freebairn_2018, lee_2010, si_2022}, and often additionally shared in formats that are difficult to decipher \cite{lee_2010, si_2022}. The value of the model evidence may also not be well understood \cite{mbachu_2024, si_2022}. Overall, these barriers are confounded by a lack of collaboration or trust between the research and political world \cite{mbachu_2024, si_2022}. 

The greatest facilitators in overcoming these barriers, included contact, collaboration, and strong relationships between policymakers and modellers \cite{oliver_2014, baker_2022, freebairn_2018, guglani_2017, alahmadi_2020}, additionally noting the importance of trust and mutual respect \cite{oliver_2014, baker_2022, freebairn_2018, si_2022}. Importantly, to promote the use of modelling evidence in decision science, policymakers noted that there should be frequent interdisciplinary exchange between the two groups, alongside early involvement \cite{baker_2022, teerawattananon_2022, guglani_2017, alahmadi_2020, si_2022}.  

Importantly, poor understanding or communication of modelling results to stakeholders can lead to significant consequences, including intentional or unintentional “misinformation, disinformation, and censorship, or, rather, public perceptions of such” \cite{levin_2024}. This may further lead to an eroding of trust in public health, institutions, or interventions \cite{levin_2024, grieve_2023}. During outbreaks, it is crucial to uphold accountability to scientific standards, consider appropriate evidence when making decisions, and remain open and transparent in communication while implementing evidence-based interventions \cite{levin_2024, grieve_2023, teerawattananon_2022}. 

In order to promote the use of modelled evidence by policymakers, modellers should ensure results are presented with consistent messaging, utilising simple, clear language, noting uncertainties, and in a lightweight format \cite{levin_2024, mbachu_2024, cairney_2017, alahmadi_2020, si_2022}. Results should be interpreted for a specific policy, using health-system generated data for models within the appropriate context \cite{mbachu_2024, cairney_2017, grieve_2023, alahmadi_2020, si_2022}, and researchers should be trained in their ability to communicate to a policy audience \cite{si_2022}. More crucially, stakeholders and modellers should be brought together in advance of outbreaks to build effective relationships and trust \cite{levin_2024, alahmadi_2020, si_2022}.

\subsection{Limitations of modelling in outbreaks}
Modelled estimates cannot substitute for a complete lack of data, but if there are several incomplete or biased data sources they may be able to help triangulate between them to get a more accurate picture. However, these estimates are no replacement for accurate surveillance and the timely collection of data, rather they act as one among a suite of tools for public health action. 
Mathematical and statistical models are not always the most appropriate method for gaining insight and can lead to misconceptions on the degree of certainty around the current epidemiological situation. 
They are also dependant on some degree of data which is sparse at the beginning of an outbreak, and may include both implicit and explicit assumptions. 
These limitations must be understood and communicated effectively to public health practitioners and policymakers \cite{Dembek2018Best, metcalf2017opportunities}. 
Ultimately, models are only as good as the data that inform them. 

\section{Conclusions}
Mathematical modelling is one facet of a multi-pronged scientific response to an outbreak of a vaccine-preventable disease. In this review, we thematically outline the important role of modelled estimates in informing outbreak response vaccination strategies and in guiding policy worldwide. We demonstrate that mathematical models can be employed to successfully quantify the impact of response timeliness, spatio-temporal heterogeneity, vaccine availability and surveillance gaps on outbreak size and in doing so, influence the design of an optimal immunisation response. While data uncertainties can be plenty, the definition of an outbreak and the pathway to outbreak detection are important factors to consider in any policy-oriented modelling exercise to measure vaccine impact. As we continue to face the threat of infectious disease outbreaks, this review emphasises that models can be used to evaluate the impact of vaccines beyond the timeline of the outbreak to help policymakers plan for population-wide healthcare needs based on available resources in the future. 

Future efforts at designing a rapid yet effective outbreak response vaccination strategy will require a holistic approach where modelling efforts are accompanied by strengthened surveillance systems, improved collaboration and communication between modellers and policy-makers as well as a contextual understanding of the pathogen, disease and demography.

\section{Declarations}

\subsection{Ethics approval and consent to participate} : NA

\subsection{Consent for publication}: NA

\subsection{Availability of data and materials}: NA

\subsection{Competing interests}: KAMG reports speaker fees from Sanofi Pasteur outside the submitted work. All other authors have no conflicts of interest to declare.

\subsection{Funding}
This work was carried out as part of the Vaccine Impact Modelling Consortium (www.vaccineimpact.org), but the views expressed are those of the authors and not necessarily those of the Consortium or its funders. The funders were given the opportunity to review this paper prior to publication, but the final decision on the content of the publication was taken by the authors. This work was supported, in whole or in part, by the Bill \& Melinda Gates Foundation, via the Vaccine Impact Modelling Consortium [Grant Number INV-034281], previously (OPP1157270 / INV-009125) and Gavi, the Vaccine Alliance. Under the grant conditions of the Foundation, a Creative Commons Attribution 4.0 Generic License has already been assigned to the Author Accepted Manuscript version that might arise from this submission.

MS, A-MH, EG, HK, J-HK, HF, MJ, AP , CLT, and KAMG received funding from Gavi, BMGF and/or the Wellcome Trust via VIMC during the course of the study. MS, A-MH, GN-G, AC, CLT, KAMG also acknowledge funding from the MRC Centre for Global Infectious Disease Analysis (reference MR/X020258/1), funded by the UK Medical Research Council (MRC). This UK funded award is carried out in the frame of the Global Health EDCTP3 Joint Undertaking. AC also acknowledges funding by the National Institute for Health and Care Research (NIHR) Health Protection Research Unit in Modelling and Health Economics, a partnership between the UK Health Security Agency, Imperial College London and LSHTM (grant code NIHR200908).

\subsection{Authors' contributions}
KAMG and CLT supervised and administered the project. MS, AMH, KAMG, CLT, RM, AP conceptualised the manuscript. MS, AMH, KAMG, GNG curated the data for the manuscript. MS, AMH, KAMG conducted the formal analysis. KAMG and CLT acquired funding. MS, AMH, KAMG , EG and GNG conducted the investigation. MJ devised the methodology for the review. MS, AMH, KAMG, CRKA, GNG and EG wrote the main manuscript. MS, AMH, KAMG, CLT, RM, CRKA, HK, HF, AP, AC, JHK, GNG, MJ reviewed and edited the manuscript. KAMG added in the visualisation and validated the manuscript. All authors reviewed and approved the final manuscript.

\subsection{Acknowledgements}

We would like to express our gratitude to Kim H Woodruff, Nick Scott, Matthew Joseph Ferrari, Xiang Li, Samaila Jackson Yaga, Helen Rees, Romesh Abeysuriya, Virginia Pitzer, Susie Cornell, Elizabeth Lee and Alex Perkins for their valuable insights and contributions during the discussions that led to the development of this manuscript. Their thoughtful feedback and perspectives were instrumental in shaping the direction of this review.

\bibliographystyle{plain}
\bibliography{refs}

\end{document}